# Granger Causality Stock Market Networks: Temporal Proximity and Preferential Attachment

Tomáš Výrost[a] – Štefan Lyócsa[b] – Eduard Baumöhl[c]

Faculty of Business Economics in Košice, University of Economics in Bratislava, Slovakia
tel.: +421 55 722 3298

[a] Department of Corporate Finance; tomas.vyrost@euke.sk; corresponding author
[b] Department of Quantitative Methods; stefan.lyocsa@euke.sk
[c] Department of Economics; eduard.baumohl@euke.sk

**Abstract**

The structure of return spillovers is examined by constructing Granger causality networks using daily closing prices of 20 developed markets from 2$^{nd}$ January 2006 to 31$^{st}$ December 2013. The data is properly aligned to take into account non-synchronous trading effects. The study of the resulting networks of over 94 sub-samples revealed three significant findings. First, after the recent financial crisis the impact of the US stock market has declined. Second, spatial probit models confirmed the role of the temporal proximity between market closing times for return spillovers, i.e. the time distance between national stock markets matters. Third, preferential attachment between stock markets exists, i.e. spillover from market *j* to market *i* is more likely if A) market *j* influences other markets other than *i*, or when B) market *i* is influenced by other markets other than *j*.

**Keywords:** stock market networks, Granger causality, emerging and frontier markets, non-synchronous trading, preferential attachment,

**Highlights:**

- Granger causality networks are constructed among 20 developed stock markets.
- A detailed procedure of handling the non-synchronicity of daily data is proposed.
- The spatial probit model is used to study the structure of the created networks.
- Relationships between markets depend on a temporal proximity of closing times.

# 1. Introduction

In empirical finance literature, one is only rarely faced with an analysis of several hundreds or thousands of relationships. However, early works of Mantegna [1] and Mantegna and Stanley [2] introduced graphs into the financial literature as a means to cope with the scale and number of complex relationships between/within economic agents. Suppose a graph $G = (V, E)$, $V \subset \mathbb{N}$, where vertices $V$ correspond to markets, and each edge $(i, j)$ from a set of edges $E$, $E \subset V \times V$, corresponds to an interaction between two markets $i$ and $j$. Such a graph represents a structure of interactions between markets. Using graph specific indicators and statistical methods, one could answer empirically or theoretically motivated questions, e.g. which markets tend to be clustered together, what type of markets tend to be on the periphery, but also why and when this happens.

Most of the network studies on financial markets study correlation based networks. Assume $N$ assets and a correlation matrix $\mathbf{C}$ of returns (with elements $\rho_{ij} \in \mathbf{C}$) with $N(N-1)$ mutual correlations $\rho_{ij}$ (excluding diagonal elements). Using suitable filtration methods, one can extract the most important correlations, which results in a much more parsimonious representation of market correlations (Mantegna [1], Coelho et al. [3]), which are in turn used to construct market graphs ready for further statistical analysis. The two dominant approaches for filtering the most important relationships are: (i) hierarchical methods and (ii) threshold methods.

Among the hierarchical methods, the most prominent representatives are minimum spanning trees (MST, for a more detailed treatment see Mantegna and Stanley [2]), and the planar maximally-filtered graph (PMFG, Tumminello et al. [4]). Numerous studies have shown that after such reductions, the vertices (asset classes) formed meaningful (usually incomplete) clusters based on industry classification or the geographical proximity of markets, e.g. Onnela et al. [5], Tumminello et al. [6], Tabak et al. [7], Lyócsa et al. [8], Bonanno et al. [9], Coelho et al. [10], Gilmore et al. [11], Eryiğit and Eryiğit [12], Song et al. [13], Mizuno et al. [14], Naylor et al. [15][1].

Networks resulting from threshold methods are much more diverse. For example, Onnela et al. [17] has suggested the asset graph, which is created by retaining $n$-largest correlations[2]. Kullmann et al. [18], Boginski et al. [19], Huang et al. [20], Tse et al. [21], Bautin et al. [22], Nobi et al. [23], Heiberger [24], Curme et al. [25] constructed networks, where for any pair of vertices, an edge is created if the corresponding correlation coefficient increases some threshold value θ, say $|\rho_{ij}| > \theta$. Sometimes, the threshold varies or is determined via statistical tests. Threshold networks were also created in Yang et al. [26] and Tu [27], where an edge was created if a standard Engle and Granger [28] test suggested a presence of a co-integration between the prices of the two assets.

---

[1] An exception is perhaps the study of Jung et al. [16], but even in this case the stocks on the Korean equity market had a tendency to cluster based on their membership in the MSCI Korea Index. This might be explained by behavioural tendencies of foreign investors, who are perhaps more trusting and therefore trade more stocks in the MSCI Korea Index compiled by an international institution than others.

The main disadvantage of the hierarchical approaches described above (MSTs and PMFGs) is that the topological constraints on these networks do not necessarily have economic or statistical rationale. On the other hand, threshold approaches need a critical value above/below which all edges are retained. Either an arbitrary value is chosen or a statistical validation is performed (e.g. Curme et al. [25], Yang et al. [26], and Tu [27]).

In this paper we use *Granger causality networks* to model the complex relationships of return spillovers between 20 developed stock markets around the world. We contribute to the existing literature in several ways. First, our construction of stock market networks is based on Granger causality testing. Second, our approach enhances the literature on threshold stock market networks by providing a sensible alternative for the choice of the threshold value. Third, we show that the role of the US market within the networks has declined over time and that the markets have become less centralized. Fourth, using the spatial probit model, we are able to confirm that the time distance between markets influences return spillovers, thus also the topology of the Granger causality networks. Even small markets, which are localized near important markets, may gain great importance in the resulting network. Fifth, we found evidence for preferential attachment between markets.

Although our approach is unique, the idea of exploiting lead-lag relationships was already used in the econophysics literature as early as in 2002 by Kullmann et al. [18], and later used in Curme et al. [25] and discussed in length by Sandoval [29]. Moreover, Granger causality networks were also already used in the finance literature of an influential paper by Billio et al. [30] and are a common tool in human brain mapping, e.g. Bullmore and Sporns [31].

---

[2] Or *n*-smallest distances from a distance matrix D, where $d_{ij} \in \mathbf{D}$, $d_{ij} = (2(1 - \rho_{ij}))^{0.5}$, see Mantegna and Stanley [2].

## 2. Data and methodology

### 2.1 Data sources

In our analysis we use daily closing prices from $N = 20$ stock market indices from four continents (Austria, Australia, Belgium, Canada, Switzerland, Germany, Spain, Finland, France, United Kingdom, Greece, Hong Kong, Ireland, Italy, Japan, Netherlands, Norway, Portugal, Sweden, and United States)[3]. Our sample starts in 2$^{nd}$ January 2006 and ends on 31$^{st}$ December 2013. Markets were selected on the basis of the availability of data and closing hours, including information on changes in closing hours (see Section 2.3). Prior to the analysis, all prices were converted into US dollars, to mimic the perspective of a US-based investor. As we are working with daily closing prices, exchange rates should have a negligible impact on the resulting time series.

### 2.2 Granger causality test

Networks created in this paper are based on the notion of Granger causality, which is a term coined by applied researchers using the principles of cross-dependence between time-series as described in the works of Granger [32], [33]. Assume, that the information set of a time series $\{x_{it}\}$ available at period $t$ is $I_{it}$. We say that $x_{it}$ is Granger causing $x_{jt}$, with respect to $I_t = I_{it} \cup I_{jt}$ if:

$$E(x_{jt}|I_{jt-1}) \neq E(x_{jt}|I_{t-1}) \tag{1}$$

In this paper we utilize Granger causality tests which are based on the cross-correlation function of standardized conditional mean returns (Cheung and Ng [34] and Hong [35]).

For each series of returns $r_t$, $t \in T$ we estimate a suitable ARMA $(p, q)$ model:

$$\begin{aligned} r_t &= \alpha + z_t \\ \left(1 - \sum_{i=1}^{p}\phi_i L^i\right) z_t &= \left(1 + \sum_{j=1}^{q}\theta_j L^j\right)\varepsilon_t \\ \varepsilon_t &= \sigma_t \eta_t, \quad \eta_t \sim iid(0,1) \end{aligned} \tag{2}$$

where $\alpha$, $\phi_i$, $\theta_j$ and $\sigma_t$ are model parameters. To account for asymmetries and long-tail properties of returns, we allow $\eta_t$ to follow a Skewed-Generalized Error Distribution. The variance $\sigma_t^2$ is modelled using a GARCH model. A standard GARCH $(r, s)$ model of Bollerslev [36] is specified as:

---
[3] According to the Dow Jones Country Classification System (as of September 2011) all these countries are considered to be developed countries.

$$\sigma_t^2 = \omega + \sum_{k=1}^{r} \alpha_k \varepsilon_{t-k}^2 + \sum_{l=1}^{s} \beta_l \sigma_{t-l}^2 \tag{3}$$

with model parameters $\omega$, $\alpha_k$ and $\beta_l$. We have considered other specifications, where the preferred variance equation was chosen from the following alternative models: AVGARCH (Taylor [37]), NGARCH (Higgins and Bera [38]), EGARCH (Nelson [39]), GJR-GARCH (Glosten et al. [40]), APARCH (Ding et al. [41]), NAGARCH (Engle and Ng [42]), TGARCH (Zakoian [43]), FGARCH (Hentschel [44]), CSGARCH (Lee and Engle [45]).

For each series, we consider a set of ARMA($p$,$q$)-GARCH($r$,$s$) models with $p, q, r, s \in \{1, \dots, 4\}$. Only models with no autocorrelation and conditional heteroskedasticity in standardized residuals are considered. For this purpose we have utilized the Peña and Rodríguez [46] test with Monte Carlo critical values (see Lin and McLeod [47]). After identifying the viable models, we have preferred the more parsimonious models, that is, the ones with the least number of estimated parameters ($p$, $q$, $r$, $s$). Where necessary, the chosen specification was finally selected using the Bayesian information criterion (Schwartz [48]).

The resulting standardized residuals ($s_{it} = \varepsilon_{it}/\sigma_{it}$) of the two given series were aligned (see Section 2.3) and used to calculate the cross-lagged correlations:

$$\hat{\rho}_{ij}(k) = \frac{\hat{C}_{ij}(k)}{\sqrt{\hat{C}_{ii}(0)\hat{C}_{jj}(0)}} \tag{4}$$

where:

$$\hat{C}_{ij}(k) = \begin{cases} \dfrac{1}{T}\sum_{t=k+1}^{T} s_{it} s_{jt-k}, & k \geq 0 \\ \dfrac{1}{T}\sum_{t=-k+1}^{T} s_{it+k} s_{jt}, & k < 0 \end{cases} \tag{5}$$

The null hypothesis of Granger non-causality ($r_j \rightarrow r_i$) is tested using the test statistic proposed by Hong [35]:

$$Q_{ij}(M) = \frac{T\sum_{k=1}^{T-1} w^2(k/M)\hat{\rho}_{ij}^2(k) - \sum_{k=1}^{T-1}(1-k/T)w^2(k/M)}{\sqrt{2\sum_{k=1}^{T-1}(1-k/T)(1-(k+1)/T)w^4(k/M)}} \tag{6}$$

Where we use the Bartlett weighting scheme:

$$w(z) = \begin{cases} 1-|z|, & |z| < 1 \\ 0, & |z| \geq 1 \end{cases} \tag{7}$$

Using the Bartlett weighting scheme $w(z)$ has some practical advantages as Bartlett has a compact support, i.e. $w(z) = 0$ for $|z| > 1$. Therefore within our empirical application only correlations $\rho(k)$, $k = 1, 2, \ldots, M - 1$ need to be calculated as $w(1) = 0$. Here $M$ (bandwidth) is the lag truncation parameter. Moreover, Hong [35] shows that the choice of $M$ and $w(z)$ does not affect the size of the test (at least when a non-uniform weighting scheme is used, e.g. Bartlett or Quadratic Spectral), while power is affected only slightly. Under the null hypothesis, $Q_{ij}(M)$ follows (asymptotically) the standardized normal distribution (it is a one-sided test). Note that (7) is calculated for a given (pre-determined) bandwidth $M$, which in our empirical application was chosen to be $M = 5$, as it corresponds to one trading week, but the results were similar with $M = 3$.

Several developed markets (e.g. in Europe) share the same closing hours. Following Lu et al. [49] we consider this by allowing for instantaneous information spillover from market $j$ to market $i$, by allowing $k = 0$ in calculating cross-lagged correlations, i.e.:

$$Q_{ij,u}(M) = \frac{T\sum_{k=0}^{T-2}w^2(k/M)\hat{\rho}_{ij}^2(k) - \sum_{k=1}^{T-1}(1-k/T)w^2(k/M)}{\sqrt{2\sum_{k=1}^{T-1}(1-k/T)(1-(k+1)/T)w^4(k/M)}} \quad (8)$$

The procedures described above are applied for rolling sub-samples of 3 months, with a drift parameter equal to 1 month, where the number of observations with each sub-sample varies slightly. This leaves us with 94 sub-samples. This approach is similar to that presented in Lu et al. [49], who used a sub-sample of 100 observations, which according to a rule of thumb of Belle [50] corresponds to a type II error of 0.05. Size of our sub-samples is around 65 observations which might lead to lower power of the test. With 380 relationships being tested, we account for a multiple-comparison problem by adjusting the significance level for each test to be $0.01/(N(N-1))$, where $N$ is the number of stock markets.

## 2.3 Return alignment

From the previous section on Granger causality it is obvious, that one needs to take into account information sets, i.e. the closing hours of national stock markets. We will call this process return alignment instead of synchronization, as for most markets, returns cannot be synchronized at all because trading simply does not end at the same time (the following procedure extends the work of Baumöhl and Výrost [51]).

Suppose we want to test for the presence of Granger causality between returns from market $i$ to market $j$ (denoted as $i \rightarrow j$). We prepare the data using the following steps:

(i)  We remove observations of calendar day *t* if for market *i* or *j* the given day was a non-trading day (holidays, technical standstills, etc.)

(ii)  We calculate continuous returns $r_{it} = \ln(P_{it}/P_{it-1})$, where $P_{it}$ denotes the daily closing price of market *i* at date *t*, over all consecutive trading days, but excluding returns over non-trading week days. Returns from Friday to Monday are considered consecutive.

(iii)  We align returns based on the closing hours at markets *i* and *j*. For example, if market *i* closes at 4:00 p.m. and market *j* at 3:00 p.m. (time-zones adjusted, e.g. UTC), we use returns from market *i* at *t*−1 to explain returns on market *j* at *t*. Similarly, if market *j* closes at 5:00 p.m, we now use returns from market *i* at *t* to explain returns on market *j* at *t*. In general, if we want to test the Granger hypothesis *i*→*j*, we want to explain returns on market *j* at time *t* using the most current past return of market *i*. If proper return alignment is not performed, either we: (a) end up with tests, where future returns are used to explain past returns (from the Granger causality point of view, it does not make sense), or (b) we are explaining returns on market *j* at time *t* using much older data on market *i*, which reduces our ability to find meaningful relationships[4].

The procedure described above needs to be performed for all tests separately (e.g. not only for *i*→*j* but also for *j*→*i* tests). It is obvious, that the time at which the closing price is determined matters. For this reason we had to take into account several additional issues necessarily related to such an analysis:

(i)  It seems that as most studies report only up-to-date closing hours, they do not take into account possible changes in closing hours. When preparing our dataset, we searched not only for up-to-date closing hours but also for historical changes in trading hours. This issue is important for the Granger causality analysis (based on the daily data of stocks around the world) but has been mostly ignored in empirical research. Besides searching through home pages of stock markets and searching on the web, we double-checked our findings by contacting all stock market exchanges in our sample[5].

(ii)  Not all countries (or even all regions of a single country) use daylight savings time. Moreover, the date of a transition from summer to winter time differs. These changes were taken into account as well.

---

[4] Sandoval's [29] correlation matrix includes such relationships. Note that although several studies use Wednesday-to-Wednesday or Friday-to-Friday returns based on closing prices of a given day to overcome the synchronization issues by using weekly returns, their approach is disputable, as they ignore the fact, that these closing prices are determined at a particular hour of a day. See the discussion in Baumöhl and Lyócsa [52], who suggest using returns between average daily prices of consecutive weeks.

[5] Markets which did not respond in the first survey were contacted again after one month.

(iii) Closing hours may be influenced by the regime of how the final, closing price is determined, e.g. the closing auction. If the closing auction was not based on the last known price (i.e. the price from the closing auction might be different from the last known price of a regular trading session) we used closing hours after the closing auctions. In some instances, the exact time the market closes is determined randomly on a day-to-day basis, where closing hours are selected within a short time window after a regular trading session. In these cases (and depending on the type of closing auction) we used the last possible closing hour.

### 2.4 Granger causality networks

The Granger causality test leads straight to the construction of a directed graph $G_t = (V, E_t)$ at time $t$, with vertex set $V \subset \mathbb{N}$ corresponding to individual indices. The set of edges $E_t \subset V \times V$ contains all edges $(i, j)$ for indices $i, j \in V$ for which $i \to j$, that is, index $i$ Granger causes index $j$ at time $t$ at a given Bonferroni adjusted significance level.

Besides the standard vertex in-degree and out-degree centrality measures, we also calculate the harmonic centrality for each vertex, which can also be used for graphs that are not connected. Following Boldi and Vigna [53]:

$$H(x) = \sum_{d(x,y)<\infty, x \neq y} \frac{1}{d(x, y)} \tag{9}$$

where $d(x,y)$ is the shortest path from vertex $x$ to vertex $y$. If no such path exists, $d(x,y) = \infty$, we set $1/d(x,y) = 0$.

The stability or resiliency of the network is considered using survival ratios as in Onnela et al. [5], which denote a ratio of surviving edges. Refer to $E_t$ as a set of edges of the Granger causality network at time $t$. One-step survival ratio at time $t$ is defined as:

$$SR(1,t) = \frac{|E_t \cap E_{t-s}|}{|E_{t-s}|} \tag{10}$$

Multi-step survival ratio at time $t$ is then:

$$SR(s,t) = \frac{|E_t \cap E_{t-1} ... \cap E_{t-s}|}{|E_{t-s}|} \tag{11}$$

where $s$ is the number of steps.

## 2.5 Spatial probit

In order to understand the structure of the created Granger causality networks, we have estimated several models explaining the creation of edges within the network. As the most interesting property is the existence of linkages, the edges in the networks become the dependent variables.

The modelling of the existence/non-existence of an edge in a network naturally leads to a logit/probit type of model, with a binary dependent variable. As we consider all possible edges within a network at the same time, some issues arise. For example, it is reasonable to assume some clustering of edges might be present: if a market would be globally dominant, we might see many edges starting within the corresponding vertex. The probability of creating an edge between any two markets might therefore depend on the nature of vertices and thus the number of their existing linkages. This dependence raises some endogeneity issues with the modelling of the edge creation – clearly, the individual edges cannot be treated as independent of each other. To remedy this problem, we estimate spatial probit models proposed by McMillen [54] and LeSage [55], which take into account the interdependence between edges (for an overview of spatial models see LaSage [56]).

To construct the model, we first define the dependent and independent variables. In our setting the variable of interest corresponds to the existence of links between the given nodes. As our sample includes $N = 20$ indices, there are $N(N-1) = 380$ possible edges for each period. We set $e_{ijt} = 1$ if $(i, j) \in E_t$, otherwise we set $e_{ijt} = 0$. We call $\mathbf{E}$ the matrix of all edge indicators $e_{ijt}$. To obtain our dependent variable (designated as $\mathbf{y}$), we first vectorise the matrix of edge indicators (by calculating $\text{vec}(\mathbf{E})$), and then exclude the elements corresponding to the diagonal of $\mathbf{E}$, as we are not interested in modelling loops – these have no economic meaning in our Granger analysis. We thus obtain a vector $\mathbf{y}$ of length $N(N-1)$.

Next we define the matrix of spatial weights. Spatial econometric models frequently use the spatial weight matrices to indicate neighbouring observations, allowing for the modelling of spatial dependence. In our case, we have to define the spatial weight matrix $\mathbf{W}$ for all potential edges in $\mathbf{y}$, thus $\mathbf{W}$ is a matrix of order $N(N-1) \times N(N-1)$. In general, for any two distinct possible edges $(i, j) \in V \times V$ and $(k, l) \in V \times V$ we set the corresponding element of $\mathbf{W}$ to 1 if the possible edges share the outgoing or incoming vertex (either $i = k$ or $j = l$)[6], 0 otherwise.

---

[6] For the purposes of estimation, we have used the row standardized version of $\mathbf{W}$ where the sum of elements in each row is equal to 1.

The spatial probit models are usually constructed in two possible ways. The spatial lag model (SAR) takes the form ([56], [57]):

$$\mathbf{y}^* = \rho \mathbf{W}\mathbf{y}^* + \mathbf{X}\boldsymbol{\beta} + \boldsymbol{\varepsilon}, \quad \boldsymbol{\varepsilon} \sim \mathbf{N}\left(\mathbf{0}, \sigma_\varepsilon^2 \mathbf{I}_{N(N-1)}\right) \tag{12}$$

Here the $\mathbf{y}^*$ represents an unobserved latent variable (just like in ordinary probit), which is linked to our variable of edge indicators $\mathbf{y}$ by:

$$y_i = \begin{cases} 1, & y_i^* \geq 0 \\ 0, & y_i^* < 0 \end{cases} \tag{13}$$

for $i = 1, 2, \ldots, N(N-1)$

As can be seen from (12), the existence of an edge is modelled by the existence of other neighbouring edges, as defined by the nonzero elements of matrix $\mathbf{W}$, as well as exogenous variables $\mathbf{X}$. The model parameters include the vector $\boldsymbol{\beta}$, as well as a scalar $\rho$, which is related to spatial autocorrelation.

The other alternative in spatial probit modelling is the so-called spatial error model (SEM), with the following specification ([56], [57]):

$$\mathbf{y}^* = \mathbf{X}\boldsymbol{\beta} + \boldsymbol{\varepsilon}, \quad \boldsymbol{\varepsilon} = \lambda \boldsymbol{\varepsilon} + \mathbf{u}, \quad \mathbf{u} \sim \mathbf{N}\left(\mathbf{0}, \sigma_u^2 \mathbf{I}_{N(N-1)}\right) \tag{14}$$

Both models were estimated by Markov Chain Monte Carlo (MCMC), using 1000 draws for each of the 94 samples. As can be seen from the specification, the SEM model only incorporates spatial effects in the error terms – however, the SAR model allows for the spatial dependence within the dependent variable. Therefore, the SAR model is more interesting in our context.

## 3. Empirical results and discussion
## 3.1 Granger causality networks among developed markets

In Figure 1 we plot two out of 94 stock market networks, one with the lowest and the second with the highest harmonic centrality of the network (sum of individual vertex centralities). Several interesting observations are visible from these networks. On the left panel we see, that due to Japan and Australia, the network is not strongly connected and that the stock market of Greece is also not reachable. Interestingly, the US stock market does not have the highest vertex out-degree (as portrayed by the size of the node), but a relatively large number of in-degrees (visible by the shade of the node). At least within this sub-sample, the US stock market is not as influential as one would expect based on the size of the market. We will turn to this issue later. The network on the right was created for the period from September 2011 to November 2011. With 263 edges (compared to 165 edges before) the network is much denser, suggesting a high degree of interconnectedness between developed markets. Note also that the nodes are more similar (more uniform distribution of out-degrees) which might suggest a global contagion in the stock markets.

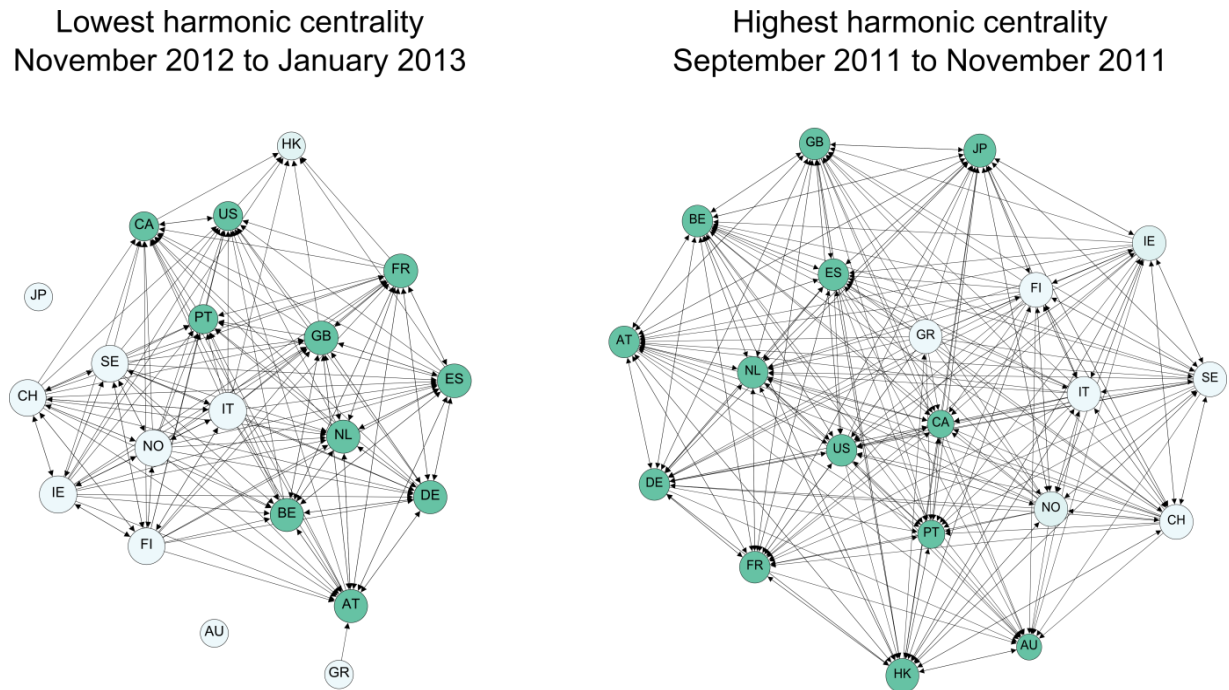

Figure 1: Directed sample networks
*Notes: Larger nodes correspond to markets with a higher out-degree, darker nodes to markets with a higher in-degree.*

## 3.2 Market out/in-degrees

Over our sample, the markets with the highest average out-degrees were situated in Europe (see Table 1). The group with highest average out-degrees of above 15 consists of Norway, Sweden, Finland, Switzerland, Ireland, and Italy. Several large markets have average out-degree above 10, i.e. Germany, France and United Kingdom. Greece had an average out-degree of 13.79 which might be due to the fact that the market is closing before other European markets (different time-zone). Within our sample, we have not confirmed the central role of the market in France as in other studies (with undirected networks), e.g. Coelho et al. [3], Gilmore et al. [11], Eryiğit and Eryiğit [12], and Sensoy et al. [58].

A surprising observation is perhaps the low average out-degree of the US stock market at just 6.83; it is the fifth smallest in our sample. A large negative linear time trend coefficient (estimate of $\beta$ in a regression: out-degrees$_t = \alpha + \beta t + \varepsilon_t$) suggests that the influence of the US market returns have declined since the beginning of 2006 (Table 1). In fact, on sub-samples up until October 2007 the average out-degree was 10.60 and even the average out-degree up until the end of 2008 in December was still 10.35, but only 4.83 after that. This suggests that the role of the US market has changed dramatically after the financial crisis.

Table 1: Descriptive statistics of Granger causality networks out/in vertex degrees

| Abb. | Market | Out-degrees | | | | In-degrees | | | | Harmonic centrality | | | |
|------|--------|------|------|--------|---|------|------|--------|---|------|------|--------|---|
| | | Mean | SD | trend | | Mean | SD | trend | | Mean | SD | trend | |
| AT | Austria | 9.83 | 1.45 | 0.015 | | 14.19 | 1.17 | -0.017 | *** | 12.95 | 1.43 | -0.005 | |
| AU | Australia | 1.31 | 3.07 | 0.000 | | 7.18 | 6.34 | 0.011 | | 3.28 | 5.02 | 0.000 | |
| BE | Belgium | 10.20 | 1.14 | 0.003 | | 14.06 | 1.24 | -0.019 | *** | 13.24 | 1.34 | -0.017 | ** |
| CA | Canada | 3.62 | 2.83 | -0.042 | * | 15.17 | 3.08 | 0.029 | | 9.18 | 3.05 | -0.063 | *** |
| CH | Switzerland | 15.63 | 1.37 | 0.016 | | 7.11 | 1.27 | -0.010 | | 16.83 | 0.90 | 0.003 | |
| DE | Germany | 10.33 | 1.22 | -0.001 | | 13.92 | 1.12 | -0.014 | ** | 13.32 | 1.34 | -0.019 | ** |
| ES | Spain | 10.33 | 1.15 | 0.001 | | 13.76 | 1.17 | -0.020 | *** | 13.33 | 1.32 | -0.019 | ** |
| FI | Finland | 15.75 | 1.41 | 0.020 | | 7.13 | 1.24 | -0.021 | ** | 16.89 | 0.94 | 0.005 | |
| FR | France | 10.46 | 1.11 | -0.001 | | 13.95 | 1.19 | -0.014 | * | 13.43 | 1.31 | -0.019 | *** |
| GB | United Kingdom | 10.32 | 1.08 | -0.002 | | 14.07 | 1.20 | -0.017 | ** | 13.32 | 1.31 | -0.019 | *** |
| GR | Greece | 13.79 | 4.64 | -0.090 | ** | 0.93 | 1.40 | -0.025 | *** | 15.76 | 3.99 | -0.071 | |
| HK | Hong Kong | 9.54 | 6.21 | -0.025 | | 6.83 | 5.75 | -0.004 | | 12.77 | 5.30 | -0.042 | |
| IE | Ireland | 15.63 | 1.38 | 0.016 | | 7.17 | 1.41 | -0.022 | ** | 16.83 | 0.95 | 0.002 | |
| IT | Italy | 15.93 | 1.42 | 0.017 | | 6.86 | 1.20 | -0.021 | *** | 16.99 | 0.94 | 0.003 | |
| JP | Japan | 4.34 | 4.87 | -0.055 | | 10.28 | 6.57 | -0.045 | | 9.70 | 4.41 | -0.060 | * |
| NL | Netherlands | 10.40 | 1.01 | -0.003 | | 14.14 | 1.13 | -0.018 | *** | 13.36 | 1.26 | -0.021 | *** |
| NO | Norway | 15.72 | 1.44 | 0.012 | | 5.69 | 2.09 | 0.044 | *** | 16.89 | 0.95 | 0.000 | |
| PT | Portugal | 2.75 | 1.74 | 0.013 | | 14.81 | 1.43 | -0.012 | | 7.44 | 3.58 | -0.002 | |
| SE | Sweden | 15.60 | 1.42 | 0.015 | | 7.00 | 1.23 | -0.016 | ** | 16.82 | 0.93 | 0.002 | |
| US | United States | 6.83 | 4.77 | -0.080 | ** | 14.05 | 3.25 | 0.045 | *** | 11.38 | 3.93 | -0.075 | *** |

*Notes: SD denotes standard deviation, trend denotes estimated β coefficients of the regression model out-degrees$_t$ = α + βt + ε$_t$, where $^{*}$, $^{**}$, $^{***}$ are used to denote statistical significance at the 10%, 5%, and 1% level, respectively. Significance is based on the HAC Newey-West standard errors estimated with automatic bandwidth selection and quadratic spectral weighting scheme as in Newey and West [59].*

In Figure 2, the plot on the left shows markets with out-degrees exceeding the 90$^{th}$ percentile for each of the sub-samples. The plot on the right marks markets with out-degrees less than the tenth percentile of that sub-sample. During and before the financial crisis, the US market was sometimes leading most markets, which changed after the crisis. In fact, the right panel in Figure 2 shows that for several occasions, the returns on the US market had the lowest impact (in terms of Granger causality) on other returns of developed markets around the world.

Asian and Australian markets in our sample are not very influential (see Table 1, Figure 2 and 3). Australia had only 1.31 out-degrees on average and Japan only 4.34, with only Hong Kong having more, with 9.54 average out-degrees. The case of the Japanese stock market might seem to be surprising as higher out-degree was expected, but the Japanese stock market is specific. It is a large market with its own important market-moving news. Moreover, it has been shown that Japanese stock market is less correlated with other markets around the world than the US market (e.g. Cappiello et al. [60], Durai and Bhaduri [61]), as Japan is a more regionally dominant stock market (Liu [62]). Regarding Hong Kong, compared to other mean out-degrees, the value of 9.54 might not seem to be a large number, but it's out-degree was highly volatile (see SD in Table 1), often reaching 16-18 out-degrees while dropping to 0-2. Note, that firms from China are also being listed on the Hong Kong stock market, which might reduce its dependence on news originating from other developed markets, particularly that of US. Not to mention that trading in Hong Kong starts later and other, non US news might already influence its returns. From this perspective, the average out-degree of 9.54 seems to be a rather high number.

The change of the role of the US market within our network is intriguing. We only hypothesize that the following factors might be responsible for this change: (i) several studies have suggested that the financial crisis increased the integration among markets (e.g. Kenourgios and Samitas [63], Syllignakis and Kouretas [64], Bekiros [65], Wang [66]), which in turn leads to an increasing importance of other developed markets around the world, (ii) after the crisis, investors might be diversifying more outside the US markets, (iii) the lower number of out-degrees of the US market might also be related to the fact, that most of the markets within our sample are in Europe, opening much later after the close of the US market[7], (iv) market moving news might be reported after trading hours (which could be used as a tool to lower the volatility of prices).

Most of the discussion above suggests that the topological properties of the stock market networks depend on the schedule of trading hours. In section 3.5 we decided to formally test this hypothesis.

From Table 1 it seems that the market's in-degree is much more evenly distributed than the market's out-degree with clear exception (see also Figure 3) only for Greece (only 0.93 in-degrees). Figure 3 also reveals (left panel) that the most influenced markets tend to be those in Australia, Canada, Japan, Portugal, and after the crisis also the stock market in the US. As these markets are probably more sensitive to global shocks, one could argue, that from the perspective of an investor these markets might not be the best choice for international portfolio diversification.

Looking at the averages reveals that, within our sample, markets with a lower in-degree tend to have a higher out-degree (see left panel of Figure 4) and vice-versa. In such cases, it might be possible, that some well situated markets (in terms of closing hours) are propagating news to other markets, while not adding too much noise from local trading (Norway, Italy, Switzerland, or perhaps Greece). For example, in times of higher global market uncertainty, when global factors are more influential, these markets might act like hubs. The right panel of Figure 4 shows how during the financial crisis the correlation between out-/in-degrees decreased, while it was higher before the crisis and at the end of our sample. Such behaviour is in line with our explanation above, as during tranquil periods such markets do not propagate much news to other markets.

---

[7] Note, that even before the crisis, the out-degree of the US stock market was around 10, still much lower than out-degrees in the European markets.

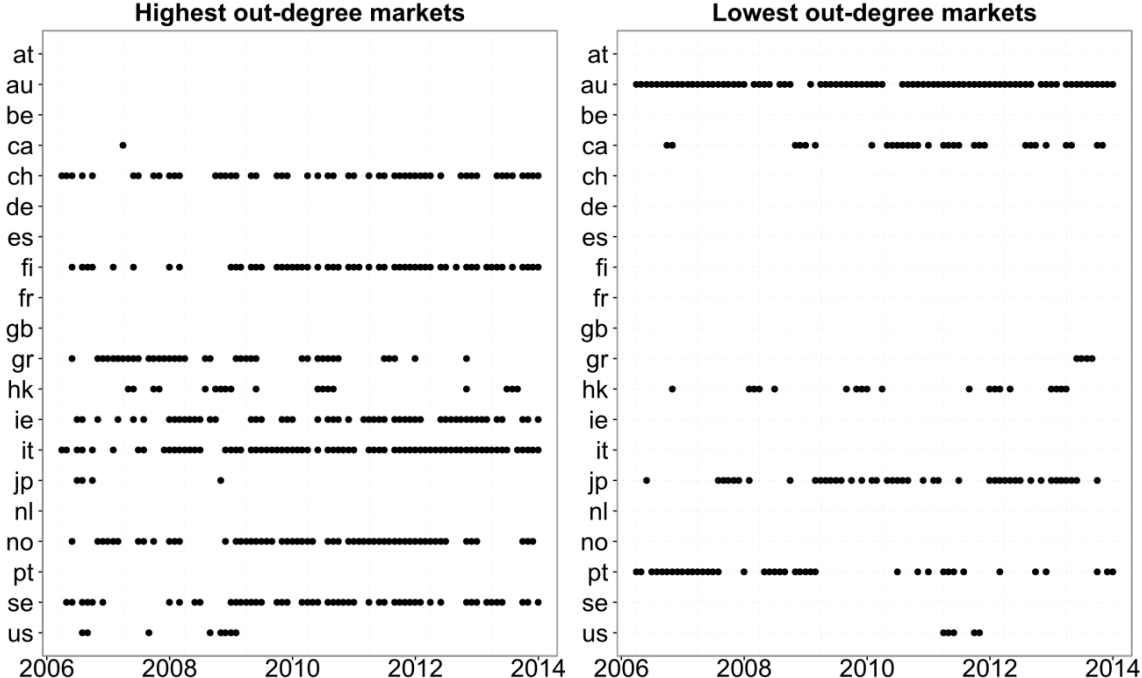

Figure 2: Occurrence of the highest (left) and the lowest (right) out-degree centralities

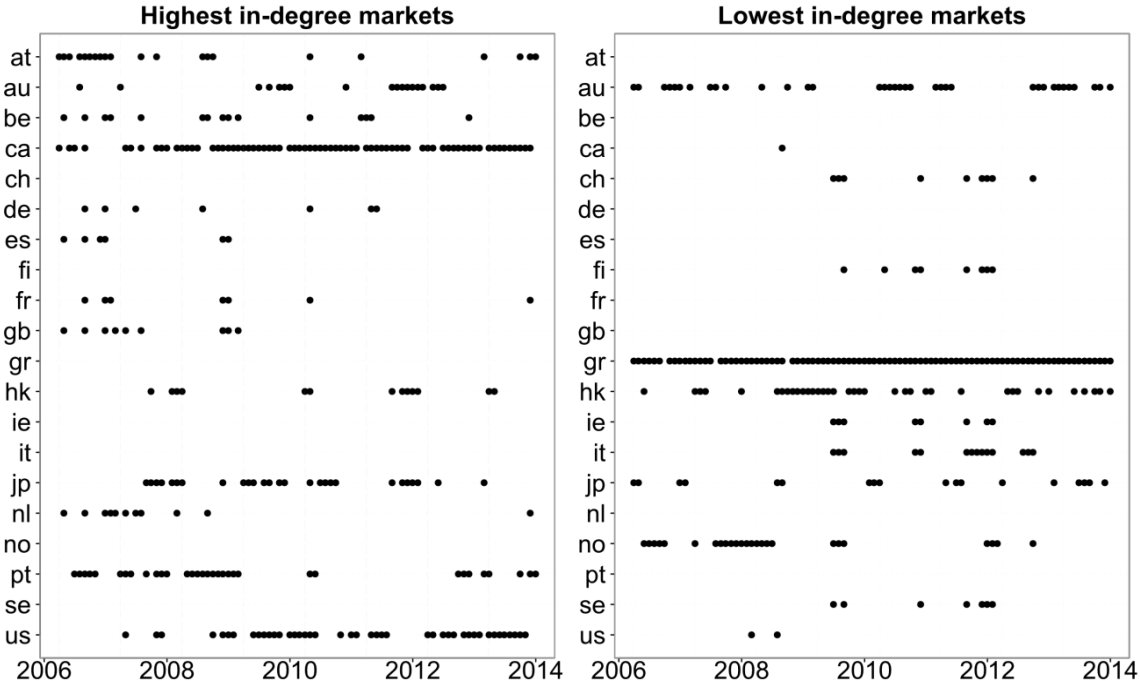

Figure 3: Occurrence of the highest (left) and the lowest (right) in-degree centralities

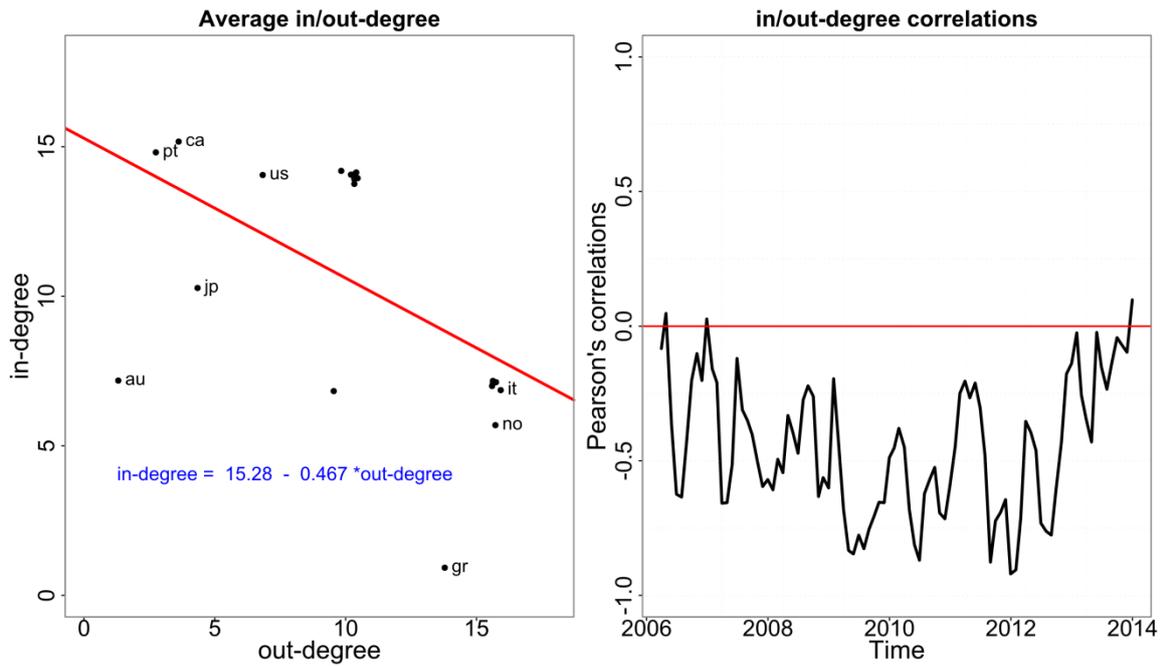

Figure 4: Out/in-degree correlations

## 3.3 Harmonic centrality

In each sub-sample and for each market we calculated harmonic centrality based on (9). Results are visualized in Figure 5 and averages over time are presented in Table 1. Markets with higher harmonic centrality are more centralized than others, probably propagating more information within the network. Almost all European markets tend to have a stable position within the network of developed stock markets, i.e. Austria, Belgium, Switzerland, Germany, Spain, Finland, France, United Kingdom, Ireland, Italy, Netherlands, Norway, and Sweden. The distribution of the market centrality is more uniform than the one observed for out-degrees. For most markets (except for Australia, Switzerland, Finland, Ireland, Italy, Norway, and Sweden) the harmonic centrality has declined over our sample period, most visibly for the markets in Canada, Greece (although not statistically significant), Japan, and the US. Overall these results suggest that the density of our networks has slightly decreased[8].

Finally, note that the centrality of the US stock market has declined after the crisis, i.e. around year 2008/2009. This further strengthens our view that, at least in terms of the influence of its returns, the role of the US stock market has declined after the crisis.

---

[8] The time-variation of network's centralization calculated as a sum of individual vertex centralities has in fact declined. Estimating a linear time trend model of centralization$_t = \alpha + \beta t + \varepsilon_t$, led to an estimate of $\beta$ coefficient of −0.418, both significant at least at the 5% significance level. Standard errors were calculated as described in the note of Table 1.

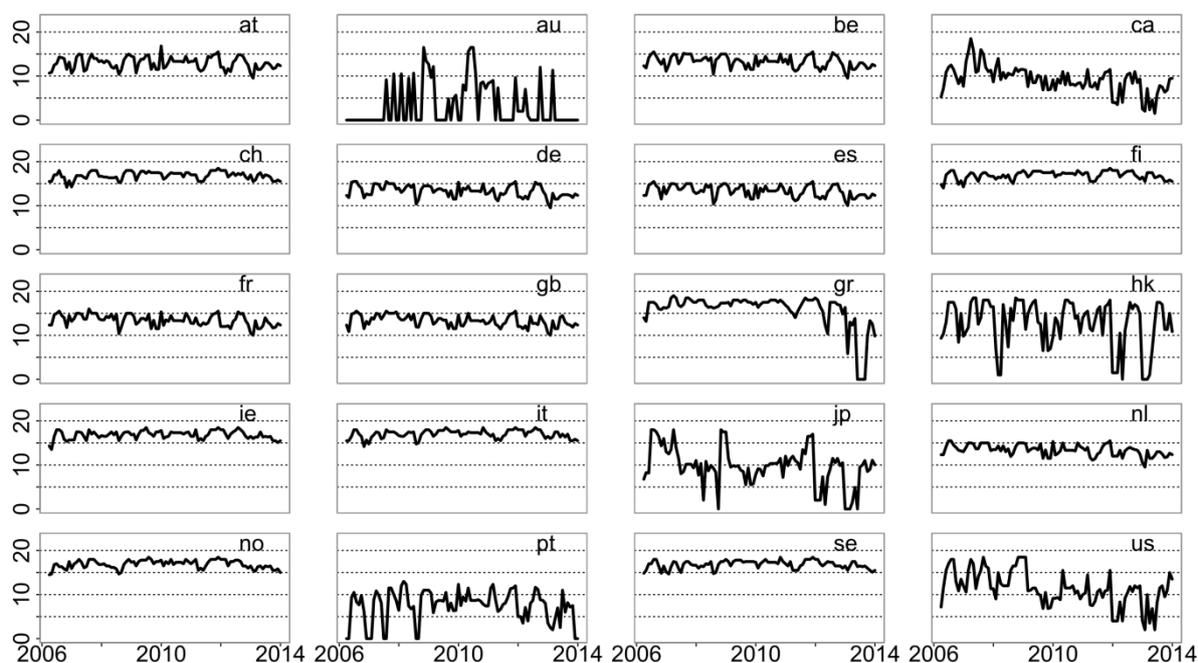

Figure 5: Market level harmonic centralities

## 3.4 Network's stability

The left panel of Figure 6 shows that the Granger causality stock market networks are rather persistent. After three steps (non-overlapping subsamples) the average survival ratio of edges is still around 85.34%. Even after 18 steps, the survival ratio was around 71.98%. This shows that many linkages among developed stock markets are rather stable over time.

On the right panel of Figure 6 we plotted the time-varying one and three step survival ratios. With three step survival ratios, one can readily observe that during the crisis, the network's resiliency was lower, but still above 60% of all edges survived even after three months.

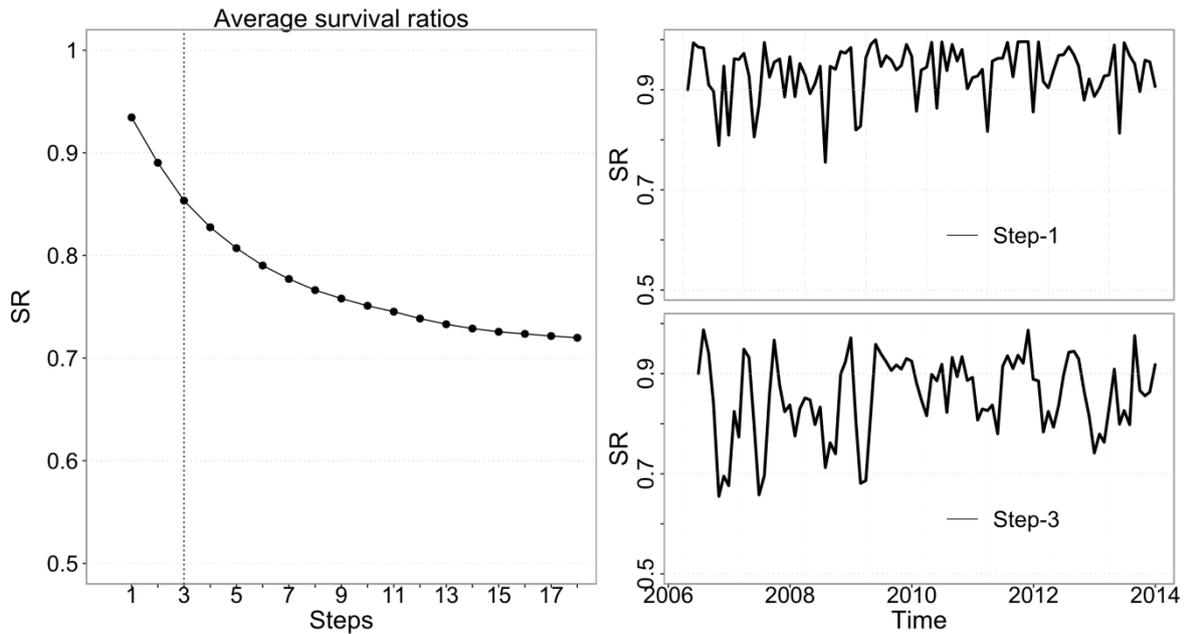

Figure 6: Average multi-step survival ratios and one/three step survival ratios

## 3.5 Structure and its determinates

The previous analysis led us to question whether the Granger causality between returns depends on the closeness of the closing hours of the markets. We hypothesize, that the structure of the stock market networks depends on closing hours, to be more specific, on their relative proximity to other closing hours. For example, if market $i$ closes just several minutes before market $j$ or a group of markets ($j \in J$) closes their trading session, it seems intuitive to expect that returns from market $i$ should Granger cause returns on markets $j$.

Each Granger causality test $i \rightarrow j$ is considered to be an observation with two possible outcomes: 1 if the null hypothesis of Granger non-causality has been rejected, or 0 otherwise. We therefore have $N(N-1)$ observations. These observations represent the dependent variable in a spatial probit model. Besides an intercept, two variables are considered as explanatory variables:

- Time (In – Out) = Closing hours on market $j$ – Closing hours on market $i$. This variable is always positive, as aligned closing hours on market $j$ precede closing prices on market $i$ (see Section 2.3 on return alignment in Granger causality tests). Note under our data alignment, this variable does not necessarily take the time difference within one calendar date − we are interested in the temporal proximity of the succession of closing hours. We assumed that the larger the time difference, the less likely it is that the two markets would be connected with an edge in a given direction. If this variable proves to be significant it suggests, that the topology of stock markets depends on the temporal proximity of markets.

- Time to US = Closing hours on market $j$ – Closing hours on the US market. This variable is also always positive as it measure how long it takes from the closing hours on the US market to the closing hours on the respective market $j$.

The spatial probit models were estimated for each sub-sample. As the results for the spatial lag model and spatial error model are very similar, we will comment only on results from the spatial lag models presented in Table 2 and Figure 7.

Table 2: Number of significant coefficients from spatial lag and spatial error probit models

|  | lag model | | | error model | | |
| --- | --- | --- | --- | --- | --- | --- |
|  | sig. at 0.10 | sig. at 0.05 | sig. at 0.01 | sig. at 0.10 | sig. at 0.05 | sig. at 0.01 |
| Intercept | 1 [1.06%] | 1 [1.06%] | 87 [92.55%] | 3 [3.19%] | 7 [7.45%] | 78 [82.98%] |
| Time (in - out) | 0 [0.00%] | 0 [0.00%] | 94 [100.00%] | 0 [0.00%] | 0 [0.00%] | 94 [100.00%] |
| Time to US | 3 [3.19%] | 14 [14.89%] | 31 [32.98%] | 5 [5.32%] | 15 [15.96%] | 26 [27.66%] |
| $\rho$ | 6 [6.38%] | 8 [8.51%] | 68 [72.34%] | 0 [0.00%] | 0 [0.00%] | 94 [100.00%] |

*Notes: The first column denotes the number of times the coefficient was significant at a specific significance level, the value in the bracket corresponds to the percentage of significant results out of the 94 periods.*

Our first observation is that the spatial lag coefficient $\rho$ was positive and significant for almost all sub-samples. This is a strong empirical evidence of some form of preferential attachment. If we test $j \rightarrow i$, then our results point to the fact that a high out-degree of market $j$ and in-degree of market $i$ increases the likelihood of creating an edge from market $j$ to market $i$ ($j \rightarrow i$).

A positive coefficient of the variable "Time to US" suggests that the further market $j$ is situated from the US market, the more likely it is that an edge from market $j$ to market $i$ will form. A negative coefficients means that markets further away from the US market tend to have a lower number of out-degrees, as the probability of a link from market $j$ to market $i$ decreases as the time distance from the US market increases. Therefore the sign and the size of the coefficient can be interpreted as a measure of the influence of the US market, with small negative values indicating higher influence. The influence of the returns on the US market seems to vary considerably, as the coefficients at the "Time to US" variable have alternating signs, i.e. for some sub-samples positive coefficients are significant while for others negative (see Figure 7).

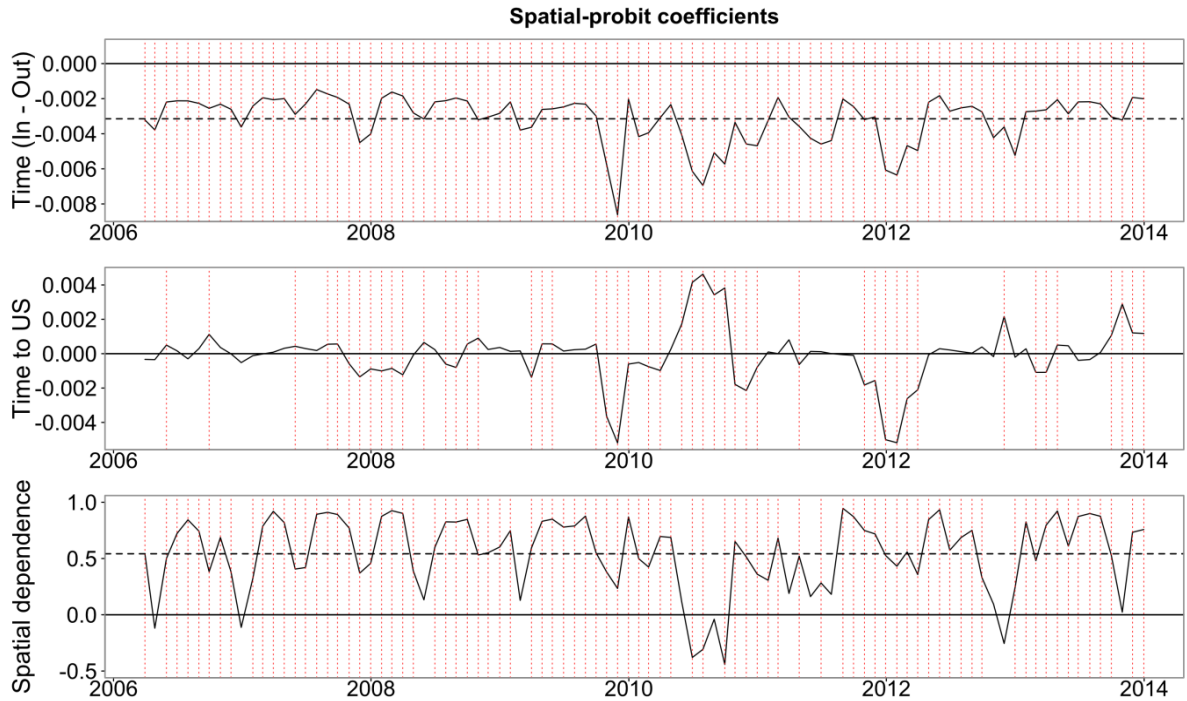

Figure 7: Time-varying spatial probit coefficients
*Notes: The red dashed lines are plotted whenever the coefficient was significant. Black dashed lines denote the average value.*

Finally, the coefficient measuring the time distance between markets $i$ and $j$ was negative (as expected) and significant for all sub-samples. It therefore seems, that although there is some variability in the effect size of the coefficient, regardless of the market conditions within our sample, the closer the market, the more likely it is that a node will be created from $j$ to $i$ (assuming we test for $j \rightarrow i$). This result clearly shows that temporal proximity matters.

## 4. Conclusion

In this paper we used a sample of daily closing prices from 20 stock markets from developed countries. Granger causality networks were constructed for 94 partially overlapping sub-samples of a length of 3 months, starting from January 2006 to December 2013. The resulting networks revealed that:

(i)  most influential returns stem from European stock markets,
(ii) before and during the financial crisis the influence of return on the US stock market was higher than after the crisis,

(iii) the most influenced returns are those on the markets in Canada, Portugal, Austria, Netherlands, United Kingdom, Belgium and after the crisis also returns on the US stock market,

(iv) networks may be considered stable, as even after a year and a half around 72% of the relationships remain present.

By estimating spatial lag and spatial error probit models we further statistically confirmed that:

(i) Market $j$, which tends to more influence other markets and market $i$, which is influenced more by other markets, also tend be more likely connected by a directional edge from market $j$ to market $i$, thus providing empirical evidence of a form of preferential attachment.

(ii) Temporal proximity between markets matters, as the closer the closing hour from market $j$ to market $i$, the more likely it is that a directional node will be created from market $j$ to market $i$.

(iii) Although we found that the role of the US market's returns have declined, it does not mean that the role of the US market is weaker. Spatial probit models revealed that there are still periods, when the temporal proximity to the US market matters.

These findings have implications for international investors, as taking into account the time distance between markets might improve decision making related to international portfolio diversification.

**Acknowledgement**

This work was supported by the Slovak Research and Development Agency under the contract No. APVV-0666-11. The usual disclaimer applies.